\documentclass[a4paper]{PoS}
\usepackage{graphicx}
\usepackage{amsmath}

\def\s0{\sigma_0}
\def\alphaEM{\alpha_{\textrm{em}}}
\def\beq{\begin{equation}}
\def\eeq{\end{equation}}
\def\bear{\begin{eqnarray}}
\def\enar{\end{eqnarray}}
\def\be{\begin{equation}}
\def\ee{\end{equation}}

\title{Drell-Yan plus jet production and BFKL evolution}

\ShortTitle{Drell-Yan plus jet production and BFKL evolution}

\author{\speaker{Krzysztof Golec-Biernat}\thanks{IFJ PAN-IV-2019-14}\\
        Institute of Nuclear Physics PAN, Krak\'ow, Poland\\
        E-mail: \email{golec@ifj.edu.pl}}

\author{Leszek Motyka\\
       Jagiellonian University, Krak\'ow, Poland\\
        E-mail: \email{Leszek.Motyka@uj.edu.pl}}

\author{Tomasz Stebel\\
      Institute of Nuclear Physics PAN, Krak\'ow, Poland \\
        E-mail: \email{Tomasz.Stebel@ifj.edu.pl}}

\abstract{Production of a forward Drell-Yan lepton pair accompanied by a jet separated by  a large rapidity interval is proposed to study  
the BFKL evolution  at the LHC.  Several observables to be measured are presented including the azimuthal angle dependence of the lepton pair 
which allows to determine Drell-Yan  structure functions. }

\FullConference{XXVII International Workshop on Deep-Inelastic Scattering and Related Subjects - DIS2019\\
		8-12 April, 2019\\
		Torino, Italy}

\begin{document}
%%%%%%%%%%%%%%%%%%%%%%%%%%%%%%%% 
\section{Introduction}

The Drell-Yan (DY)  lepton pair production  \cite{Lam:1978pu} is one of the most important processes 
which allow to study the QCD structure of the colliding hadrons.   Of particular importance
in this presentation are the high-energy QCD effects described by the Balitsky-Fadin-Kuraev-Lipatov (BFKL) 
evolution scheme \cite{Lipatov:1996ts} which is complementary to the commonly used collinear resummation schemes. A classical  
process to study the  BFKL evolution in hadronic collisions, proposed by Mueller and Navelet (MN)  \cite{Mueller:1986ey}, is a production of   two jets with similar transverse momenta but with a large rapidity interval $\Delta Y$ between them. In particular, it was proposed  to look at the azimuthal  decorrelation in  the MN jets  \cite{DelDuca:1993mn,Stirling:1994he}. Such studies were performed experimentally at the Fermilab  \cite{Abachi:1996et,Abbott:1999ai} 
and the LHC \cite{Aad:2014pua, Khachatryan:2016udy} and analyzed theoretically in
\cite{Ducloue:2013bva, Caporale:2014gpa} using the full NLO jet impact factors  and the NLL BFKL kernels.  

One of the MN jets can be replaced by a hadron \cite{Bolognino:2019ouc}.
We propose  to replace it by a forward Drell-Yan lepton pair which has 
several advantages. (i) The experimental precision of DY measurements is usually very high. 
(ii) This process offers a broader range of parameters which may be scanned like the lepton pair mass $M$ or  its  transverse momentum ${q}_\perp$.  
(iii) The lepton pair angular distribution allows to determine  the DY structure functions 
\cite{Brzeminski:2016lwh} which show sensitivity to the underlying BFKL dynamics.
(iv) Particularly interesting is  the Lam-Tung combination of the DY structure functions \cite{Lam:1980uc} which is sensitive to partons' transverse momenta.

The forthcoming  presentation is  based on the paper \cite{Golec-Biernat:2018kem} in which more details are provided.

%%%%%%%%%%%%%%%%%%%%%%%%%%%%%%%%
\section{Kinematics and cross sections}

\begin{figure}[tb]
\centerline{%
\includegraphics[width=0.35\textwidth]{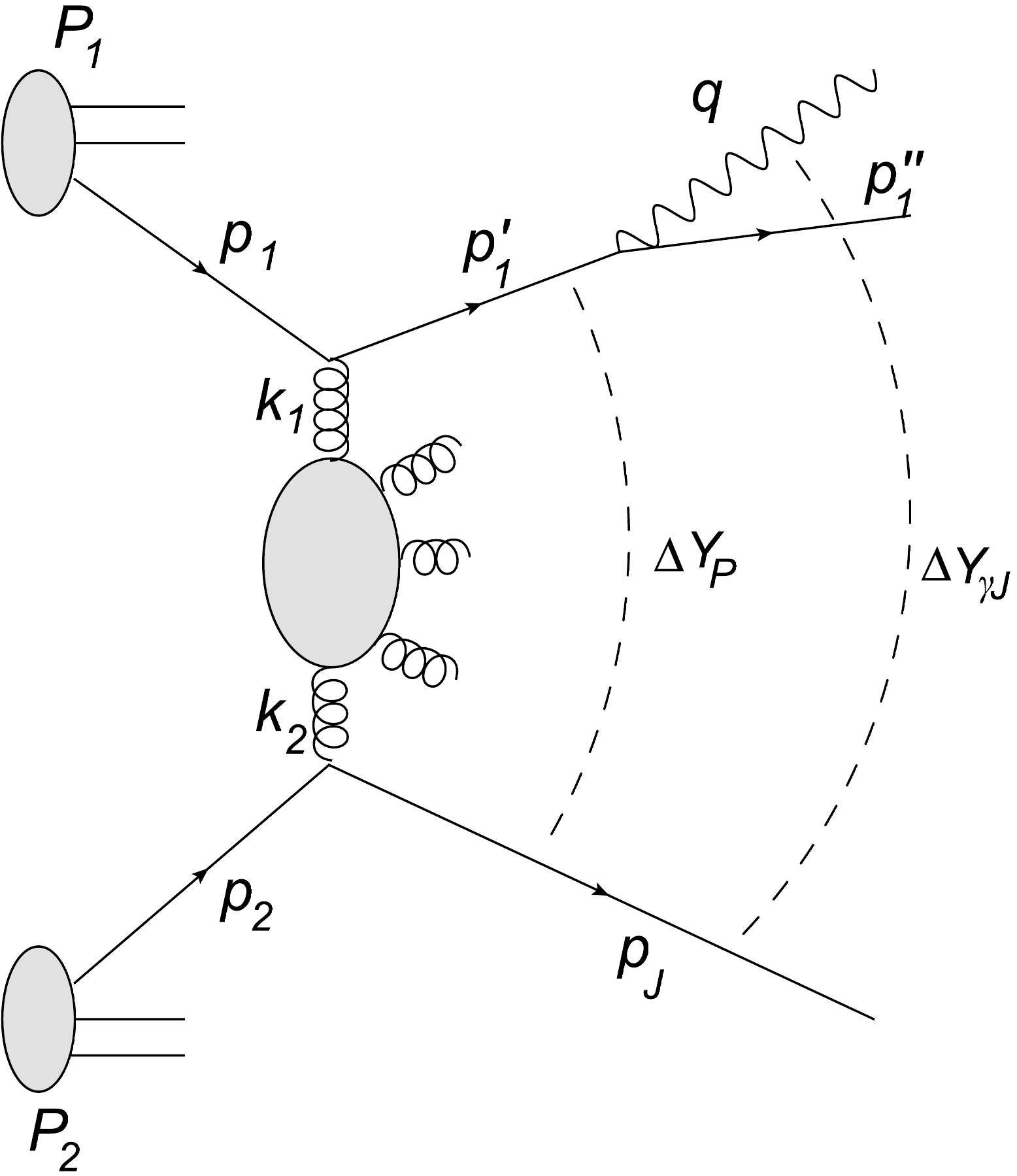}}
\caption{Kinematic variables relevant for the DY+jet production}
\label{Fig:fig1}
\end{figure}

In Fig.~\ref{Fig:fig1} we show the relevant kinematic variables for the DY pair plus jet production with the BFKL ladder exchange.
The most important is  the large rapidity interval $\Delta Y_{\gamma J}$ between the forward DY boson and the backward jet  to be measured experimentally. The rapidity distance $\Delta Y_P$ is purely theoretical since it is an argument of the BFKL kernel, related to $\Delta Y_{\gamma J}$ through  kinematics, see formulae (2.9) and (2.10) in \cite{Golec-Biernat:2018kem}.  The DY+jet  cross section reads
\begin{eqnarray}
\label{sigAsWcomb_DYj}
\frac{d\sigma^{\textrm{DY+j}}}{ d\Pi \,d\Omega}  &=& (1-\cos ^2 \theta)  \frac{d\sigma^{(L)}}{d\Pi}+ (1+\cos ^2 \theta)  \frac{d\sigma^{(T)}}{d\Pi} \nonumber \\
&+& (\sin^2\theta \cos 2\phi)\frac{d\sigma^{(TT)}}{d\Pi}+ (\sin2\theta \cos \phi) \frac{d\sigma^{(LT)}}{d\Pi},
\end{eqnarray}
where $(\theta, \phi)$ are lepton spherical angles in the helicty frame, $\Pi=(M^2, \vec{q}_\perp, \vec{p}_{J\perp}, \Delta Y_{\gamma J})$ are variables to be measured
and the DY+j structure functions  for four polarizations, $\lambda=L,T,LT,TT$, 
are given by
\begin{align}\nonumber
\label{sig_master_formula}
\frac{d\sigma^{(\lambda)}}{ d\Pi} & =
\frac{4 \alphaEM^2\alpha_s^2}{(2\pi)^4} \,\frac{1}{M^2 p_{J\perp}^2}
\int_0^1 dx_1 \int_0^1 d x_2\,\theta(1-z)\,f_{q}(x_1,\mu) f_{\textrm{eff}}(x_2,\mu) 
\\
&\times \int \frac{d^2 k_{1\perp}}{ k_{1\perp}^2 } \, \Phi^{(\lambda)} ( \vec q_\perp, \vec k_{1\perp}, z)\,
K_{BFKL}(\vec{k}_{1\perp}, \vec{k}_{2\perp}=-\vec p_{J\perp}, \Delta Y_{P}),
\end{align}
where  $\Phi^{(\lambda)}$  are the DY impact factors and $K_{BFKL}$ is the BFKL kernel  given by the Fourier decomposition
in the azimuthal angle $\phi$ between $\vec k_{1\perp}$ and $\vec{k}_{2\perp}$
\be
K_{BFKL}(\vec k_{1\perp},\vec{k}_{2\perp},\Delta Y_P) =\frac{2}{(2\pi)^2|\vec k_{1\perp}||\vec k_{2\perp}|}
 \Big( I_0(\Delta Y_P)+ \sum_{m=1}^{\infty} 2\cos(m \phi) I_m(\Delta Y_P) \Big)
\label{Fourier_exp_K}
\ee
 More details on the BFKL are given in  \cite{Golec-Biernat:2018kem}. We only mention here that in our analysis $K_{BFKL}$ is given in the leading 
 logarithmic (LL) order  with important  part of the NLL corrections which are taken into account in the form of the kinematic constraint.
 
 In the forthcoming presentation we will show the results for the helicity-inclusive cross section (\ref{sigAsWcomb_DYj})
 integrated  over the full spherical angle $\Omega$ of the lepton pair,
\beq
\frac{d\sigma^{\textrm{DY+j}}}{d\Pi } = \frac{16\pi}{3}\left( \frac{d\sigma^{(T)}}{d\Pi} + \frac{1}{2} \frac{d\sigma^{(L)}}{d\Pi} \right).
\label{hel_incl_xsect}
\eeq

%%%%%%%%%%%%%%%%%%%%%%%%%%%%%%%%
\section{Azimuthal angle dependence}

\begin{figure}[tb]
\centerline{%
\includegraphics[width=.45\textwidth]{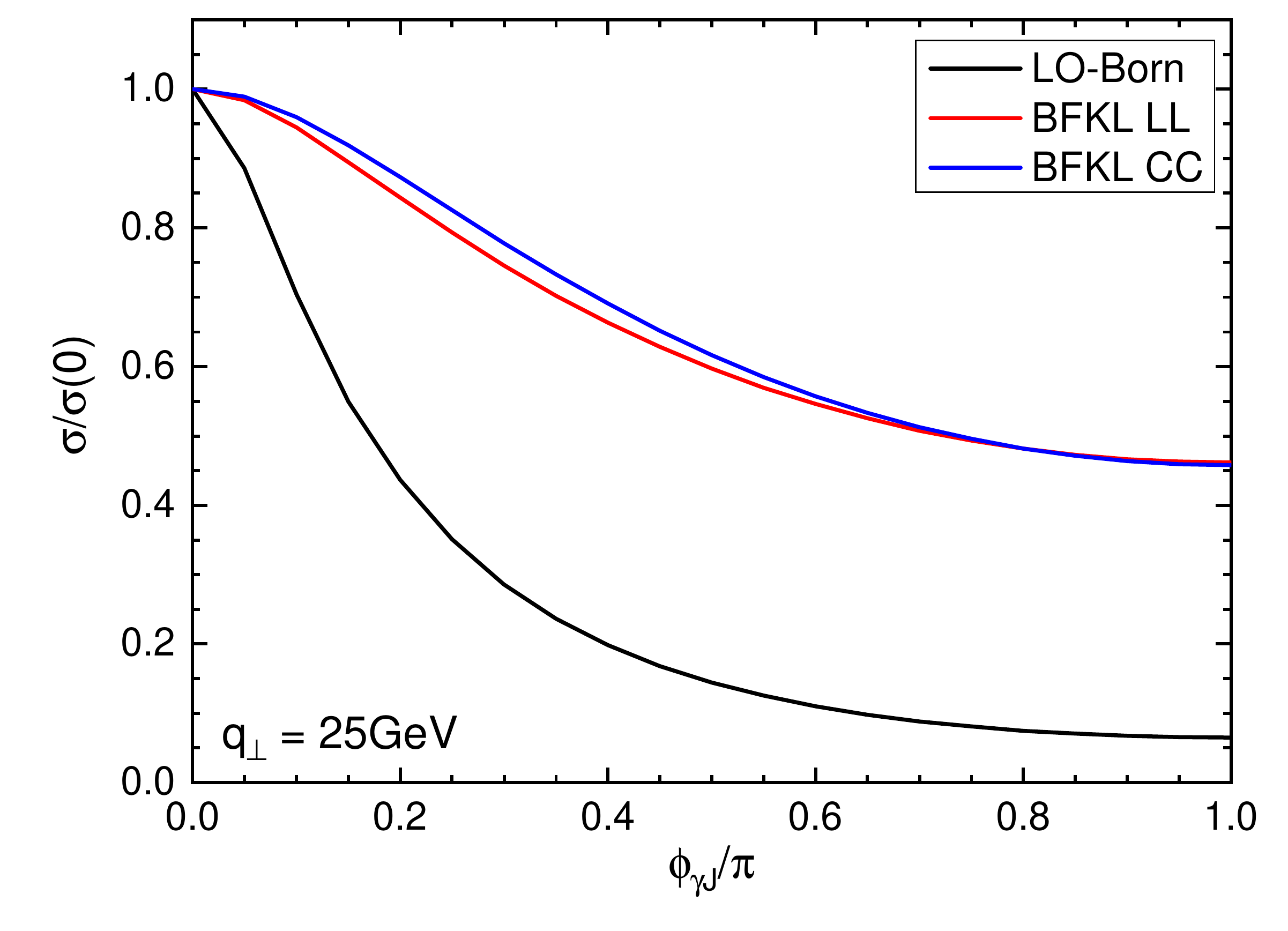}
\includegraphics[width=.45\textwidth]{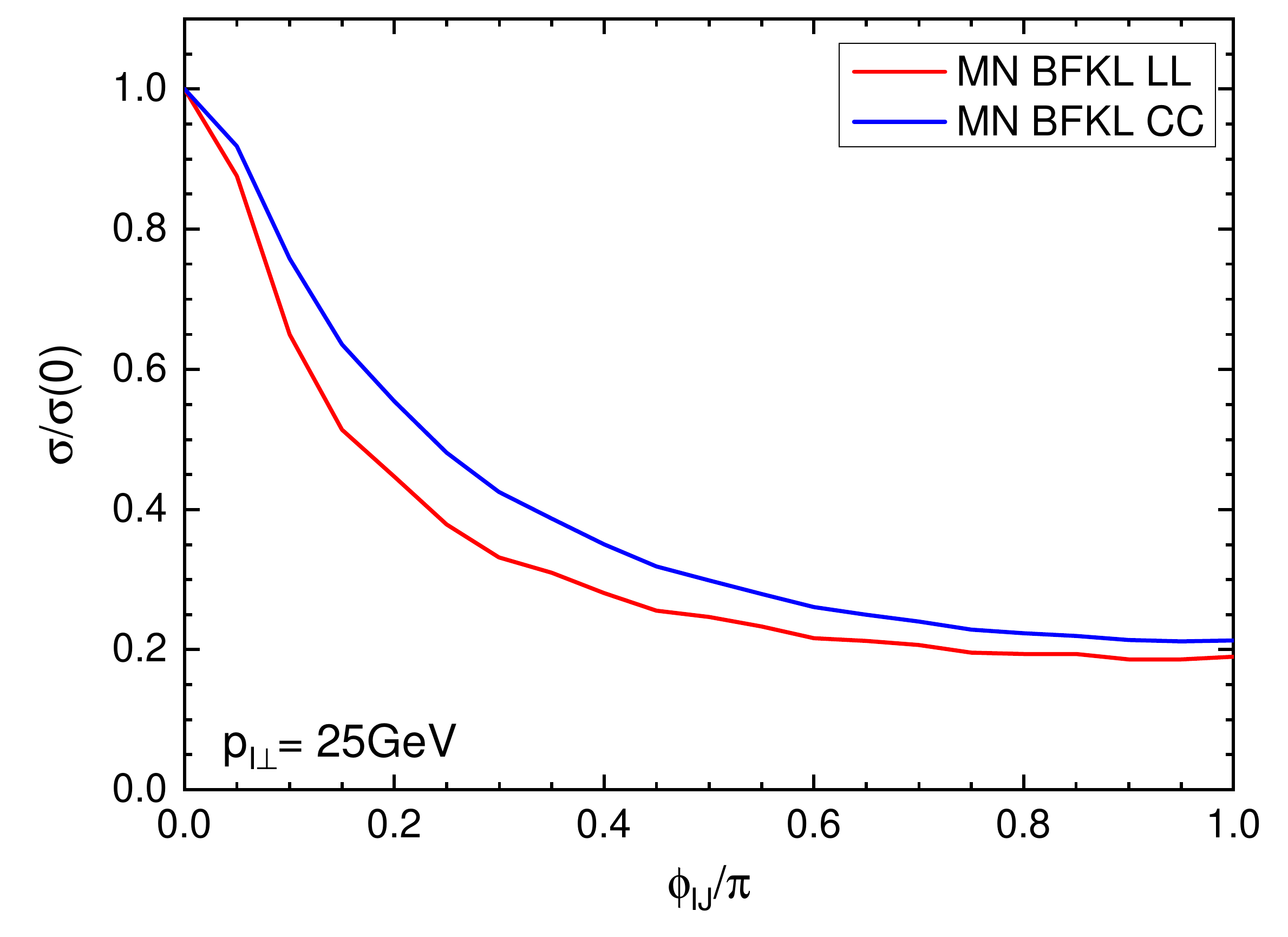}}
\caption{The azimuthal angle dependence of the normalized helicity cross sections for the DY+jet (right) 
and MN jets (left) processes for  $q_\perp=25\,{\rm GeV}$, $p_{J\perp}=30\,{\rm GeV}$, $M=35\,{\rm GeV}$
and $\Delta Y_{\gamma J}=\Delta Y_{IJ}=7$.
}
\label{Fig:fig2}
\end{figure}

\begin{figure}[tb]
\centerline{%
\includegraphics[width=.45\textwidth]{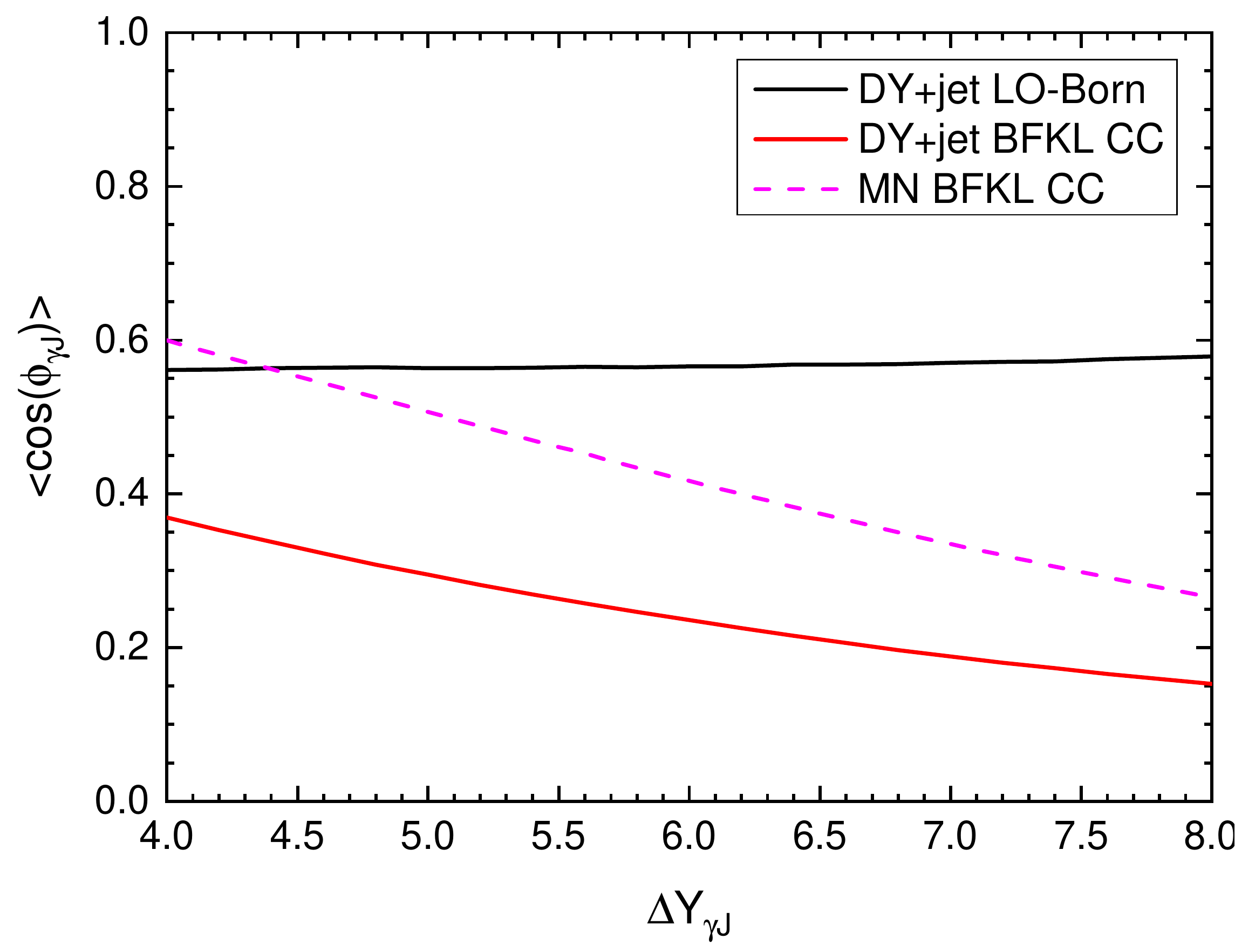}
\includegraphics[width=.45\textwidth]{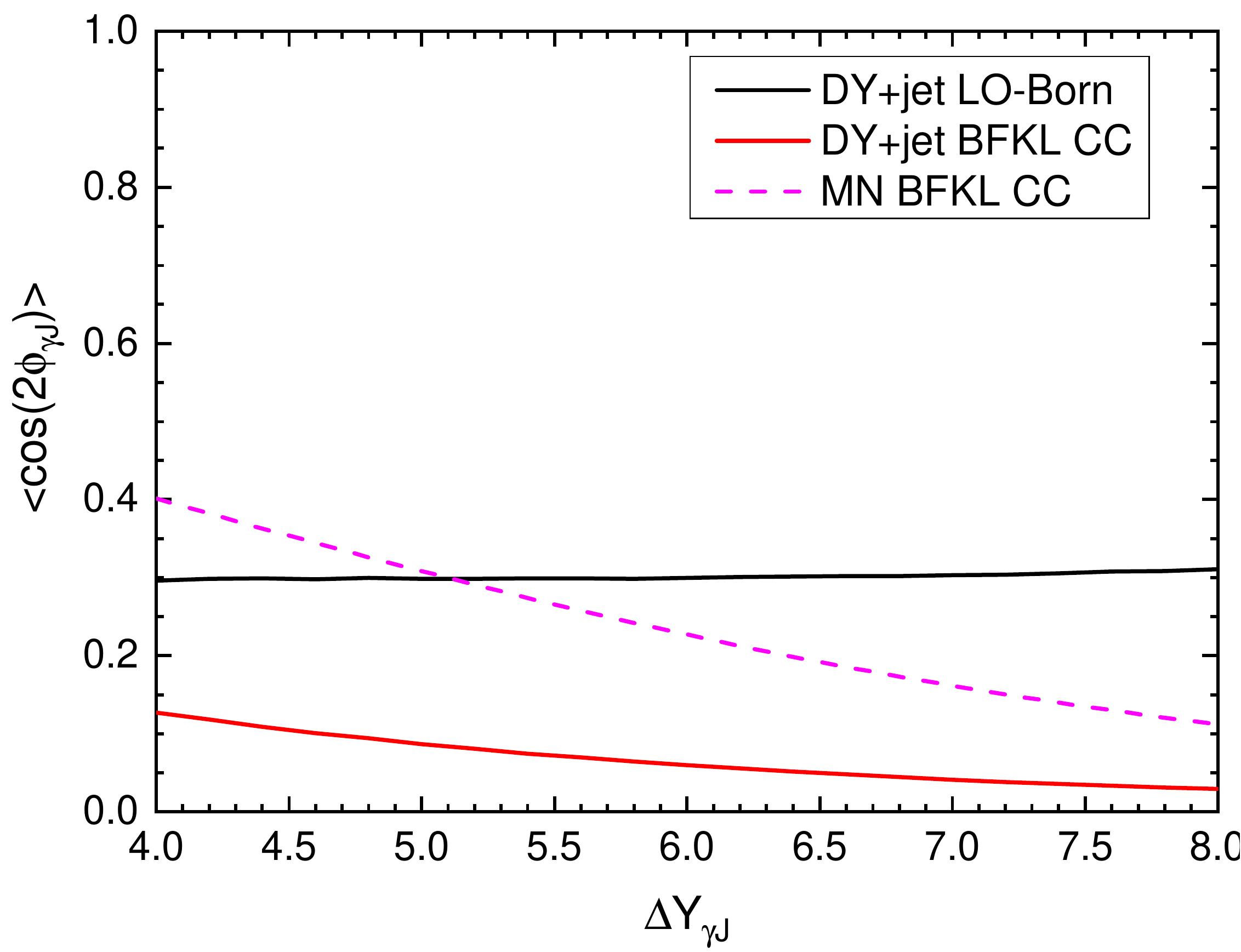} }
\caption{The mean cosine $\langle \cos (n\phi_{\gamma J}) \rangle$ as a function of rapidity difference $\Delta Y_{\gamma J}$ for $n=1$ (left) and $n=2$ (right) for the DY+jet (solid lines) and MN jets (dashed lines) processes 
with $q_\perp=p_{I\perp}=25\,{\rm GeV}$,  $p_{J\perp}=30\,{\rm GeV}$ and $M=35\,{\rm GeV}$.
}
\label{Fig:fig3}
\end{figure}

The first quantity to study is the  dependence of (\ref{hel_incl_xsect}) on the azimuthal angle
$\phi_{\gamma J}$ between the photon transverse momentum $\vec q_\perp$ and jet momentum $\vec p_{J\perp}$,
see Fig.~\ref{Fig:fig2} where the normalized formula (\ref{hel_incl_xsect})  is shown as a function
of $\phi_{\gamma J}$. The BFKL  effects in the DY+jet case (left) lead to stronger decorrelation in the azimuthal angle in comparison to the LO-Born (two gluon) exchange and also to the MN jet  case (right).

In Fig.~\ref{Fig:fig3} we show the comparison between the DY+jet (solid lines) and MN jet (dashed lines) processes in terms of the mean cosine
\beq
{\langle \cos (n\phi_{\gamma J})\rangle}=\frac{\int_0^{2\pi} d\phi_{\gamma J} \ \frac{d\sigma^{\rm DY+j}}{dM d\Delta Y_{\gamma J} dq_\perp \, d p_{J \perp}d\phi_{\gamma J}}\cos( n\phi_{\gamma J}) }
{\int_0^{2\pi} d\phi_{\gamma J} \ \frac{d\sigma^{\rm DY+j}}{dM d\Delta Y_{\gamma J} dq_\perp \, d p_{J \perp}d\phi_{\gamma J}} }
\eeq
We study this quantity  as a function of the rapidity difference $\Delta Y_{\gamma J}$ for   $n=1$ and $n=2$ .
In both cases, we see stronger decorrelation for the DY+jet production that for the MN jet case. 
Note that  the mean cosine values equal one for the  MN jet process in the Born approximation,
when both jets have the same transverse momentum, which gives the strongest possible correlation.

%%%%%%%%%%%%%%%%%%%%%%%%%%%%%%%%
\section{Angular coefficients}

In the inclusive DY process, it is useful to define normalized structure functions. We follow this approach and define
 the following coefficients for the DY+jet process
 \begin{equation}
A_0 = \frac{d\sigma^{(L)}  }{ d\sigma^{(T)} + d\sigma^{(L)} /2 }, \ \ \ \ A_1 = \frac{d\sigma^{(LT)}  }{ d\sigma^{(T)} +  d\sigma^{(L)}/2  } ,  \ \ \ \ A_2 =  \frac{2d\sigma^{(TT)}  }{ d\sigma^{(T)} +  d\sigma^{(L)}/2  }\, .
\label{A_coeff_def}
\end{equation}
Of particular interest is to study the Lam-Tung relation
\beq
 d\sigma^{(L)}- 2 d\sigma^{(TT)}=0~~~~~~~~ \textrm{ or } ~~~~~~~~A_0-A_2=0.
\label{Lam-Tung_combination}
\eeq
which is  valid at the LO and NLO for the DY $qg$ channel in the  leading twist collinear approximation \cite{Lam:1978pu,Lam:1980uc}

\begin{figure}[tb]
\centerline{%
\includegraphics[width=.46\textwidth]{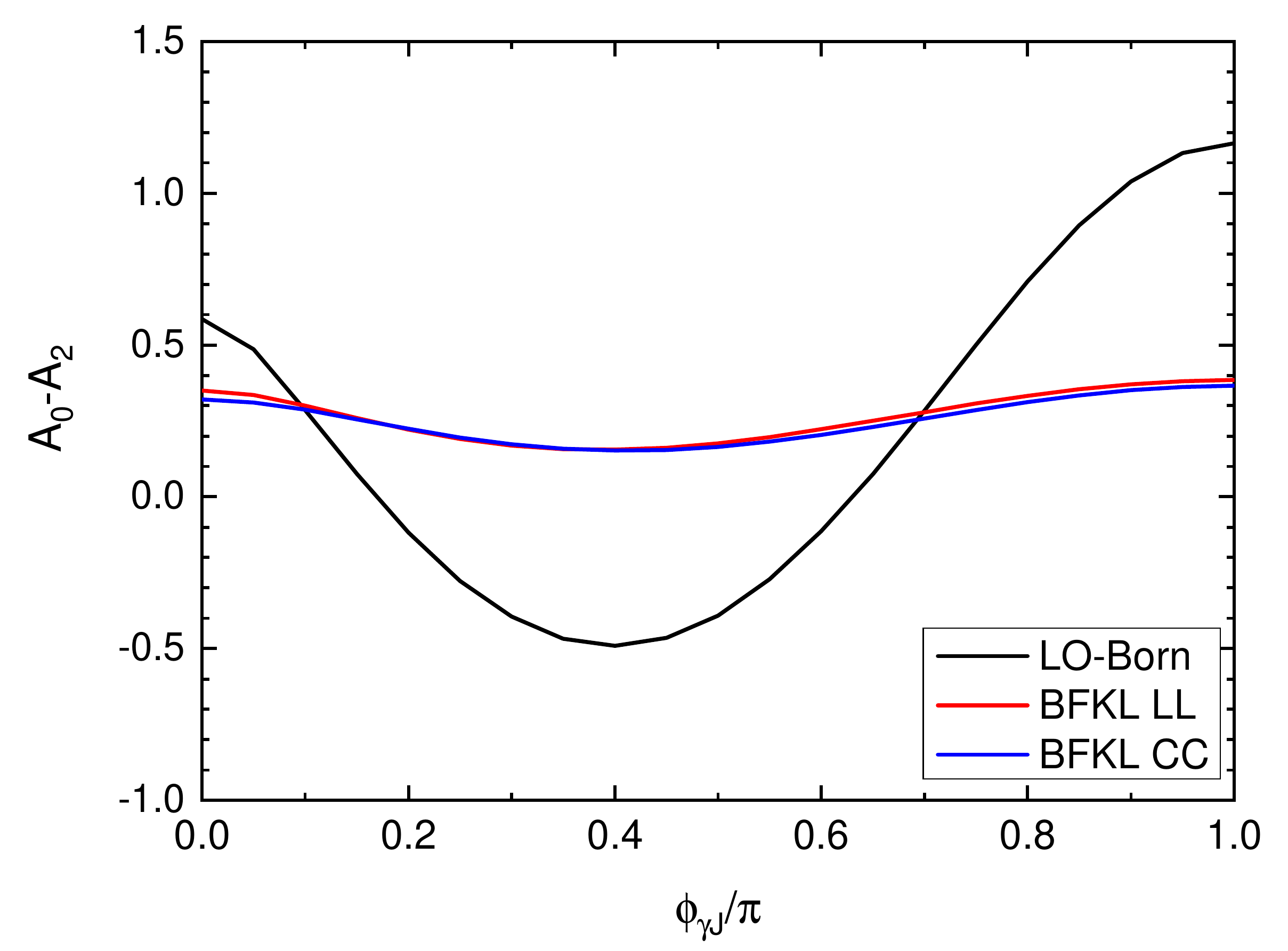}
\includegraphics[width=.46\textwidth]{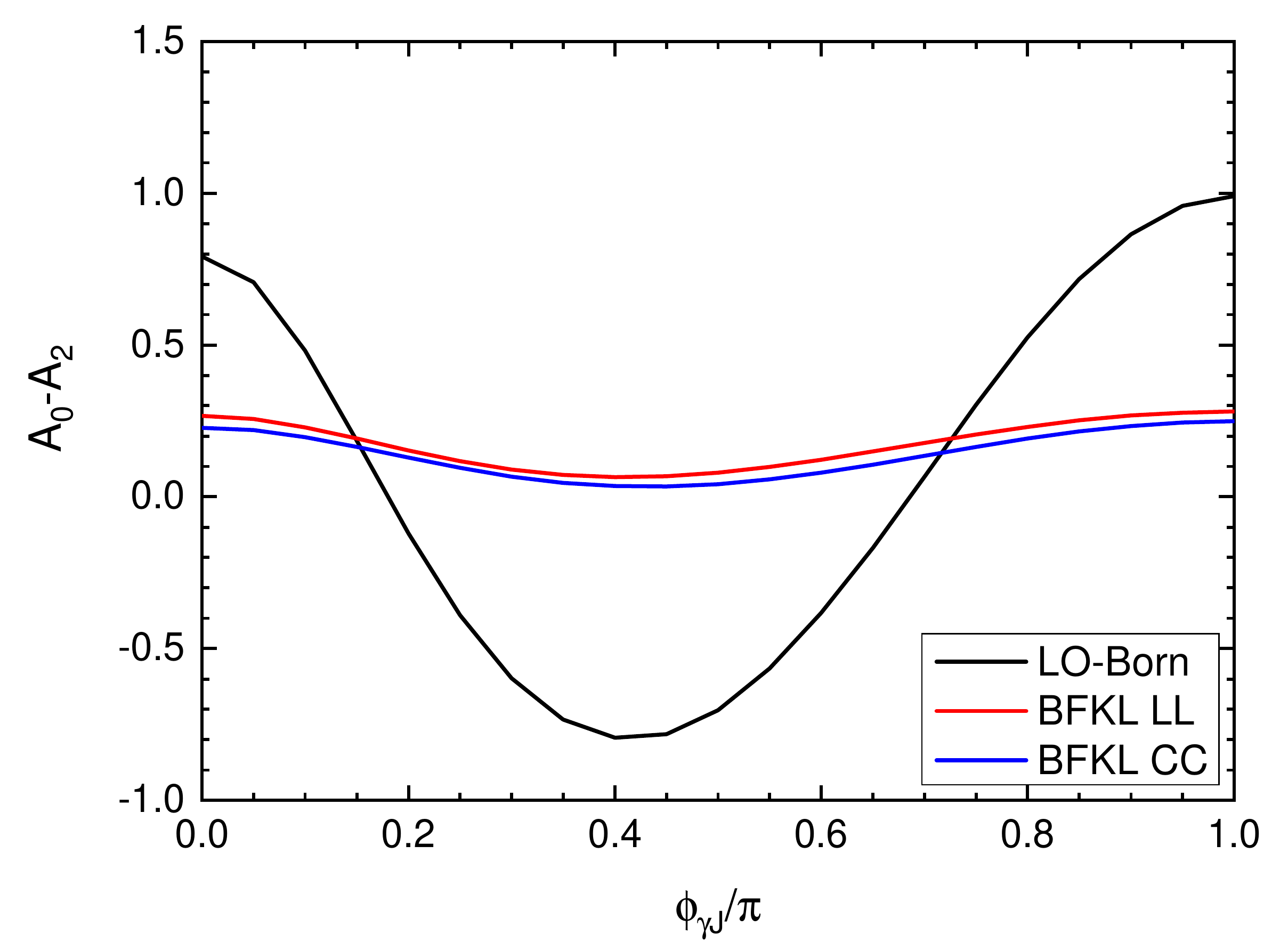}}
\caption{The Lam-Tung  difference  of angular  coefficients, $A_0-A_2$, as a function the azimuthal photon-jet angle
 $\phi_{\gamma J}$ for $q_\perp=25\,{\rm GeV}$ (left) and $q_\perp=60\,{\rm GeV}$ (right) 
 while $p_{J\perp}=30\,{\rm GeV}$, $\Delta Y_{\gamma J}=7$ and $M=35\,{\rm GeV}$.}
\label{Fig:fig4}
\end{figure}

As it was shown in \cite{Motyka:2016lta}, the combination $A_0-A_2$ is sensitive to partons' transverse momenta.
In Fig.~\ref{Fig:fig4} we show this combination as a function the photon-jet azimuthal angle $\phi_{\gamma J}$.
We see a dramatic difference between the full BFKL result, which is almost independent of the angle, and the LO-Born approximation (two gluon exchange) in which we find 
strong dependence on the angle $\phi_{\gamma J}$.
A similar pattern can be found for the coefficients $A_0,A_1$ and $A_2$, separately. This shows  that 
for leptons' angular coefficients, the decorrelation coming from the BFKL emission is almost complete.

%%%%%%%%%%%%%%%%%%%%%%%%%%%%%%%%%
\section{Conclusions}

W proposed a new process to study the BFKL evolution in high energy hadronic collisions  at the LHC -- the Drell-Yan  plus jet production. In this process, the DY photon with large rapidity difference with respect to the backward jet should be tagged through the lepton pair.
The presented numerical results show a significant angular decorrelation with respect to the Born approximation for the BFKL kernel. The found decorrelation is also stronger than for the Mueller-Navelet jets due to more complicated final state with one more particle, being the  DY photon. We also presented numerical results on the angular coefficients of the DY lepton pair which provide an additional experimental opportunity to test the effect of the BFKL evolution in the proposed process. Of  particular interest is  the Lam-Tung relation
which is sensitive to partons' transverse momenta.

The future studies should include the full NLL BFKL kernel as well as the NLO photon plus jet impact factors. Although the first element is known, the NLO impact
factors have not been computed yet.

%%%%%%%%%%%%%%%%%%%%%%%%%%%%%%%%%%
\vskip 5mm
\centerline{\bf Acknowledgments}
\vskip 2mm

 This work was  supported by the National Science Center, Poland, Grants No. 2015/17/B/ST2/01838, DEC-2014/13/B/ST2/02486 and 2017/27/B/ST2/02755. 

%%%%%%%%%%%%%%%%%%%%%%%%%%%%%%%%%%
\bibliographystyle{JHEP}
\bibliography{mybib}

%\begin{thebibliography}{99}
%\bibitem{...}
%\end{thebibliography}

\end{document}